\documentclass[12pt]{article}
\begin{document}
\title{{\bf Our Place in a Vast Universe}
\thanks{Alberta-Thy-19-07, arXiv:0801.0245, to be published in Melville
Y.~Stewart, ed., {\em Science and Religion in Dialogue} (Blackwell
Publishing Inc., Oxford), and in Melville Y.~Stewart and Fu Youde, eds.,
{\em Science and Religion: Current Dialogue} (Peking University Press,
Beijing, in Chinese), from a series of lectures sponsored by the
Templeton Foundation and given at Shandong University in Jinan, China,
autumn 2007; see also arXiv:0801.0246 and arXiv:0801.0247.}}
\author{
Don N. Page
\thanks{Internet address:
don@phys.ualberta.ca}
\\
Institute for Theoretical Physics\\
Department of Physics, University of Alberta\\
Room 238 CEB, 11322 -- 89 Avenue\\
Edmonton, Alberta, Canada T6G 2G7
}
\date{(2008 January 4)}

\maketitle
\large
\begin{abstract}
\baselineskip 20 pt

	Scientists have measured that what we can see of space is about a
billion billion billion billion billion billion billion billion billion
($10^{81}$) times the volume of an average human.  Inflationary theory
suggests that the entirety of space is vastly larger.  Quantum theory
suggests that there are very many different copies of space of the same
basic kind as ours (same laws of physics).  String theory further
suggests that there may be many different kinds of space.  This whole
collection of googolplexes of galaxies within each of googolplexes of
different spaces within each of googols of kinds of space makes up an
enormously vast universe or multiverse.  Human beings seem to be an
incredibly small part of this universe in terms of physical size.  Yet in
other ways, we may still be a very significant part of our vast universe.

\end{abstract}
\normalsize

\baselineskip 19.6 pt

\newpage

    The monotheism of Judaism, Christianity, and Islam asserts that the
universe was created by God.  For example, the first verse in the Bible,
Genesis 1:1, says, ``In the beginning God created the heavens and the
earth'' \cite{Bible}.  In the New Testament, the third verse of the
Gospel of John, John 1:3, says, ``All things were made through Him [Jesus
Christ as the Son of God], and without Him nothing was made that was
made.''

    Astronomers, cosmologists, physicists, and other scientists who study
the universe continue to discover that the universe is larger than
previously thought.  Human beings seem to be an incredibly small part of
the universe in terms of physical size.  Yet in other ways, we still
believe that we are a very significant part.

    Ancient people knew that the universe was much larger than humans. 
For example, Psalm 8 in the Bible, composed perhaps around 3,000 years
ago, refers to this qualitative knowledge:  ``When I consider Your
heavens, the work of Your fingers, the moon and the stars, which You have
ordained, what is man that You are mindful of him, and the son of man
that You visit him?''

    The Greeks were perhaps the first to try to measure how much larger
than humans the earth and other astronomical objects are.  Around 2,200
years ago, Eratosthenes measured the size of the earth to within a few
percent of the correct value.  About the same time, Aristarchus of Samos
made observations of the size of the shadow of the earth on the moon
during an eclipse to deduce the distance to the moon.  He had correct
geometric reasoning but made poor measurements, getting the distance from
the earth to the moon as 20 times the radius of the earth, whereas the
actual average distance is about 60 times.

     Aristarchus also measured the angle between the sun and the moon
when the moon was half-illuminated (quarter moon), as well as other
angles, to estimate the distance from the earth to the sun to be nearly
400 times the radius of the earth.  This estimate was actually too small
by a factor of about 60 because of errors in measuring the angles, some
of which were too small or too near a right angle to be measured
accurately at that time.  However, if one combined these measurements
with the viewpoint of some of the ancient Greeks who contended that the
radius of the universe was the distance from the earth to the sun, one
could have deduced that the universe was over a billion times larger than
an average human.

    Aristarchus thought that the stars were actually much further away
than the sun, which led him to believe the universe was considerably
larger than the distance from the earth to the sun.  Although none of
Aristarchus' writings on this matter have survived, Archimedes wrote that
Aristarchus believed the sun and stars were at rest and that it was the
earth that revolved around the sun.  (This reference makes Aristarchus
the first person to propose the heliocentric theory, that the sun is at
the center of the solar system.)

    In order to avoid observable changes in angles between the directions
to the stars as the earth moved around the sun (stellar parallax),
Aristarchus had to assume that the distances to the stars were much
larger than the distance from the sun to the earth.  But most people
accepted neither this heliocentric view nor the belief that the stars
were much farther away than the sun, until Copernicus proposed similar
ideas again in the 16th century.

    Part of the objection to Aristarchus' heliocentric view may have been
an assumption that as the home of humans, the earth ought to be the
center of the universe.  However, a larger reason for assuming that the
earth was at the center seems to have been the argument of Aristotle (who
lived about 100 years before Aristarchus), that as something heavy, the
earth would have settled at the center of the universe.  Furthermore, it
might have seemed incredible that the stars were so far away that their
changes in direction would be unnoticeable from the earth if it really
did revolve around the sun.

    Aristotle did not conceive of the earth as the source of the gravity
around it, as Newton did 2000 years later, but rather assumed that what
we now call gravity would pull things toward the center of the whole
(what we would now call the universe) \cite{Danielson}.  Aristotle argued
that this pull toward the center is why the earth settled there, the
effect rather than the cause of things falling.  Although this
Aristotelian concept of gravity has proved to be wrong regarding the
actual gravitational field surrounding the earth, something resembling it
would occur even in Einstein's theory of gravity, general relativity, if
there were a sufficiently large negative cosmological constant.  This
occurs in the hypothetical anti-de Sitter space-time that is a favorite
toy model of many gravitational theorists today, though not as an
accurate model of our universe.

    By the Middle Ages, whatever exalted view of the central position of
the earth some of the ancients may have had was generally replaced by the
view of the central position of the earth as mundane, located at what
Galileo called the ``dump heap of the filth and dregs of the universe''
\cite{Danielson}.  Copernicans' praise of the sun's central location in
the solar system is now misinterpreted as a degradation of the earth to a
demoted position away from the center.  However, at the time, the praise
of the sun's new central position was an attempt to restore the lofty
status of the sun after it had fallen from a more exalted position. 

    After Aristarchus tried (and inevitably failed) to find the true size
of the solar system, it took a long time before an accurate size was
discovered.  The size of the earth could be  accurately measured to a
certain degree by the ancient Greeks because people could walk a
sufficient fraction of the circumference around it to measure the
difference in the upward direction (say relative to the direction of the
rays of the sun at high noon).  Then the relative size of the moon could
be measured by comparing it with the shape of the shadow of the earth on
the moon during a lunar eclipse.  From the size of the moon in the sky
(its angular diameter of about half a degree), one could deduce the
distance to the moon. However, since the sun and planets are considerably
further away than the moon, their relative distances could not be
measured until centuries later.

    The first precise measurements of the size of the solar system were
made during the transits of Venus of 1761 and 1769, when Venus passed
directly in front of the sun and could be seen as a small black disk
covering a tiny part of the sun.  The time it takes for Venus to cross
the observed disk of the sun depends on how far from the center it
crossed, and this depends slightly on the viewing position on the earth. 
(The rotation of the earth also affects the transit time.)  Therefore, by
taking precise timing measurements of the transit durations from
different locations on earth, one could deduce the distance to Venus and
to the sun in terms of the known distances between the locations on the
earth.  (The ratios of the distances to Venus and to the sun at various
times, though not their actual values, could be deduced centuries earlier
from the angles between the directions to the sun and to Venus at various
times.)

\newpage

    Many countries cooperated in sending expeditions to distant parts of
the earth to take these measurements of the 1761 and 1769 transits of
Venus.  Wars and bad weather hampered many attempts, such as the one made
by the unfortunate Guillaume Le Gentil of France \cite{Hogg}.  In 1761 he
could not land at Pondicherry, a French colony in India, because the
British had seized it, and he could not make his measurements from his
ship that was tossing about at sea.  He stayed eight years to make
measurements of the 1769 transit (the last transit before 1874) and this
time was able to set up his equipment on Pondicherry, which was restored
to France by then.  But after a month of clear weather, the sky turned
cloudy on the morning of the transit, and he again saw nothing.  He
nearly went insane but gained enough strength to return to France, which
took another two years.  After being away for nearly twelve years in his
fruitless mission to help measure the size of the solar system, Le Gentil
finally got back home to find that his ``widowed'' wife had remarried and
his possessions had gone to his heirs.

    Once the size of the solar system had been determined, the next
prodigious step was to measure the distances to the stars.  For a
sufficiently nearby star, this could be determined by measuring the
change in the direction from the earth to the star as the earth revolved
around the sun.  The first successful measurement of stellar parallax was
done by Friedrich Bessel in 1838, for the star 61 Cygni, a little more
than 10 light years away (about 700,000 times as far away from earth as
the sun).  Since then, the stellar parallaxes of over 100,000 other stars
have been measured, notable by Hipparcos (HIgh Precision PARallax
Collecting Satellite).

    To measure the distances to stars and galaxies that are farther away,
one must use other methods, such as the apparent brightness of stars or
galaxies for which there is independent evidence of how bright they would
be at a given distance (their intrinsic brightness or absolute
luminosity).  One obtains what is sometimes called a whole ``ladder'' of
cosmic distance scales, with a sequence of ``rungs'' or classes of
objects used to determine ever-greater distances.  That is, the distances
to nearby objects of one class are measured by one method (e.g., by
stellar parallax), and the same class of objects (say with apparently the
same absolute luminosity) is used to calibrate the next class of objects
at greater distances, which is in turn used to calibrate another class of
objects at still greater distances.

    There are also some methods, such as measuring the angles, times of
variability, and velocities of multiple images of objects focused by
gravitational lensing, that allow one to get direct measures of some
distances without using other rungs of the ladder.  (The velocities of
distant stars and galaxies can be determined by measuring how much the
wavelengths of the light they emit is shifted, a ``redshift'' toward
longer or redder wavelengths for objects moving away from us. 
Gravitational lensing is the bending of light by a mass, say of a galaxy,
that is close to the line of sight from a more distant object, the source
of the light. In some cases the light can be bent to reach us in two or
more routes from the more distant source.)

    Planets are grouped in solar systems.  Stars like the sun are grouped
in galaxies containing about one hundred billion stars, with our galaxy
being called the Milky Way, or simply the Galaxy.  Galaxies themselves
are generally arranged in groups of less than 50 galaxies, clusters of
50-1000 galaxies, and superclusters of many thousands of galaxies.

    It has been found that very distant galaxies are at distances nearly
proportional to their velocities away from us, the proportionality
constant being roughly the time since the galaxies would have been on top
of each other at the beginning of the universe.  Therefore, measuring
both the distances and the velocities of distant galaxies can tell us the
age of the universe.  For most of my academic career this age had an
uncertainty factor of nearly 2, but in recent years it has been
determined much more precisely to be roughly 13.7 billion years, likely
between 13.56 and 13.86 billion years \cite{Spergel}.

    Since we cannot see beyond the distance light has traveled since the
beginning of the universe, we cannot see further than about 14 billion
light years away.  (Actually, what we see as nearly 14 billion light
years away has been moving away from us since the light we now see left
it; one may estimate that today it might be about 50 billion light years
away, but of course we cannot see that distance yet.)

    This size of the observable universe, say 13.7 billion light years,
is nearly a million billion ($10^{15}$) times the average distance
between the earth and the sun (499 light seconds), and almost a hundred
million billion billion ($10^{26}$) times the height of a human.  If one
includes the recession of the most distant observed galaxies up to their
unobserved present distance of about 50 billion light years, one can say
that the universe we have observed now has a volume more than a billion
billion billion billion billion billion billion billion billion
($10^{81}$) times the volume of an average human.

    However, this is not the end regarding our thoughts on the size of
the universe.  The furthest distances we can see within the universe are
limited by the age of the universe, since light can travel only 13.7
billion light years if the age of the universe is only 13.7 billion
years.  But what we see at that distance shows no sign of being near an
edge or being an end to space; consequently, it appears likely that the
universe extends far beyond what we can see.  (There is also no sign that
the universe is curled up into a finite volume, as the surface of the
earth is curled up into a finite area in one lower dimension.)

    How much further the universe might extend of course we cannot
directly observe, but we can ask how far it might extend.  Basically, no
one has thought of any convincing limit that would prevent the entire
universe from being infinite.  (It often seems as if the universe would
be easier to understand if it were finite, but that is not persuasive
enough to imply that the universe must be finite.)  Even if the universe
is finite, it might be enormously larger than what we can see of it.

    Indeed, in the past few decades, a theory called inflation
\cite{Linde,Guth,Vilenkin} has been developed to explain several of the
mysteries of the universe, and it generally predicts that the universe is
much, much larger than what we can see.  One of these mysteries is the
large-scale homogeneity of the universe, the fact that at the largest
distance we can see, averaging over many superclusters of galaxies, the
universe appears to be statistically the same everywhere.  (There do not
seem to be significant structures much larger than superclusters, or
larger than about 1\% of the furthest distance we can see.)  The universe
is also highly isotropic, with the superclusters of galaxies
statistically nearly the same in all directions.  A third mystery is the
fact that the universe has expanded to become very large, and yet gravity
is still important.

    Inflation gives partial explanations of these mysteries by
postulating that the early universe expanded exponentially to become
enormously larger than it originally was.  This expansion would smooth
out whatever lumps there might have been (within some limits) and hence
make the present universe highly homogeneous and isotropic at the largest
distances.  It would also tend to lead to a balance between expansion and
gravity, so that the universe would not have already recollapsed as it
otherwise might have, and it would not have thinned out so much that
gravity would have become unimportant.

    Of course, we may note that within superclusters of galaxies, the
universe is far from homogeneous and isotropic, and gravitational
collapse has occurred to form planets, stars, and black holes. Although
inflation was not originally designed to explain the departures from the
overall approximate homogeneity and isotropy that it successfully
explained, it was a bonus that with a simple starting point, inflation
could partially explain the observed inhomogeneities and anisotropies of
our universe, which of course are essential for our existence.  The idea
is that although inflation apparently smoothed out the early universe to
a very high degree, it could not make it completely smooth, because of
what are called quantum fluctuations.

    Quantum fluctuations result in the Heisenberg uncertainty principle,
which contends that one cannot have precise values for both the location
and the momentum (mass-energy times velocity) of any object. Normally for
objects of everyday experience that are much larger than atoms, these
quantum fluctuations are too small to be noticed, though some of the
fuzziness and tiny ringlike appearance of floaters in the eye is due to
the diffraction of light, which might be viewed as one form of quantum
fluctuations.  So one might expect that quantum fluctuations would be
negligible for astronomically larger objects, such as planets, stars,
galaxies, clusters, and superclusters of galaxies.

    On the other hand, the universe is expanding, so in the past it was
much smaller.  The entire universe we can now see was apparently once
smaller than the width of a hair.  And if inflation is right, it was even
astronomically smaller than that.  For such a tiny universe, smaller than
the size of a present-day atom, quantum fluctuations can be significant.

    One can calculate that quantum fluctuations in the early universe
would lead to variations in the density of ordinary matter that formed
later, and these density variations would then clump because of their
gravitational attraction to form the planets, stars, galaxies, clusters,
and superclusters we see today.  Therefore, it appears that inflation,
acting on the inevitable quantum fluctuations, can lead to the structured
universe we see today.  Indeed, our very existence depends upon this
structure that apparently can be explained by inflation that amplifies
initially tiny quantum fluctuations.

    Although we are not absolutely certain that inflation really occurred,
it does offer quite a lot of partial explanations for what we observe. 
Therefore, it is interesting to see what other predictions inflation
makes.

    One of inflation's most important prediction is that the universe is
most likely to be enormously larger than what we can see, which is
limited by the limited present age of the universe.  This is not just
saying that the universe is likely to be billions of times larger than
what we can see.  Rather, suppose that we wrote a sequence of billions to
tell how much larger the universe is.  (Above we gave a sequence of nine
billions multiplied together, $10^{81}$, or one with 81 zeros after it,
to denote the number of humans who could fit into the volume of the part
of the universe today that we have seen.)  Not enough ink would exist in
the world to write down the number of billions multiplied together to
give the size of the universe. The size of the universe might be even
enormously greater than a googol, $10^{100}$, 1 followed by 100 zeros,
and perhaps even greater than a googolplex, $10^{10^{100}}$, 1 followed
by a googol of zeros, which would require a googol of digits to write
down as an ordinary integer.

    As a googol is itself far larger than the number of elementary
particles in the part of the universe we can now see, a googolplex could
not be written down in the ordinary way (as a sequence of digits), even
if one could write each digit with only a single elementary particle from
all those within sight.  And inflation tends to predict that the whole
universe is even larger than a googolplex.

    To put it another way, with only ten digits we could in principle
assign a different telephone number for every human now on earth.  But
with each particle from the whole observable part of the universe taken
to be a digit, we could not count the size of the total universe,
according to many inflationary theories.  Thus inflation suggests that
the entire universe is enormously larger than what we can see of it, so
much larger that it boggles the imagination.

    Furthermore, one of the most prominent modern ways to understand
quantum theory and its fluctuations (the Everett ``many worlds'' theory
\cite{many-worlds}) implies that there may be an exorbitant number of
universes, perhaps as many as the staggeringly large number of particles
within the entirety of our universe if inflation is correct.  So we think
that not only is our universe very, very large, but also there are very,
very many of them.

    Many experts believe that string/M theory, which is the leading
contender to be the physics theory of everything, predicts that there are
many different kinds of universes \cite{Susskind}.  The law of physics in
each of them would be different.  There might be one relatively simple
overarching set of laws for the collection of all universes, but in each
kind of universe, the apparent laws would differ, being merely ``bylaws''
for that kind of universe.

    In many kinds of universes the ``constants of physics'' that lead to
the structures of molecules we know would be different, so that these
molecules would be replaced by different ones.  In many other kinds of
universes, there would be no molecules at all. In many others it is
believed that even spacetime would not exist, though what would exist in
its place is still rather murky in the minds of the theorists trying to
understand string/M theory.

    Some estimates are that it would require perhaps 500 digits just to
list the different ``laws of physics'' of the distinctly qualitatively
different kinds of universes within string/M theory \cite{DouKac}.  And
remember, each of these different kinds of universes can have
googolplexes of particular universes (Everett worlds), and many of these
particular universes can be so large as to have googolplexes of
particles, planets, stars, galaxies, clusters, and superclusters of
galaxies within them.

    Suppose it took 500 digits to say which kind of universe one is in, a
googol of digits to say which of the Everett ``many worlds'' or universes
of that kind one is in, and another googol of digits to say where within
one of those inflated enormous universes one is.  Then it would take 500
plus two googols to specify where we are.  This would make us an
incredibly tiny part of the entire collection of universes, which is
sometimes called the multiverse.

    One might seek an analogy of our place in the multiverse in terms of
human experience within all the humans on the earth.  Then one could take
the 500 or so digits that specifies which kind of universe to be
analogous to a specification of the genome of a person.  However, listing
the genetic code of a human, in the form of listing the sequence of about
3 billion DNA base pairs, requires specifying about 2 billion digits,
many more digits than the 500 digits that some think would be needed to
specify the kind of universe in string/M theory.  Thus in this regard
string/M theory would apparently be much simpler than human genetics,
though at present it is perhaps even less well understood.

    The googol of digits that specifies which of the Everett many-worlds
describes our universe could be taken to be analogous to a specification
of which person one is out of all those with the same genes.  For people
living on the earth at the same time, in most cases one would not need
further specification beyond their genes, although identical twins or
other multiplets that originated from a single fertilized egg would.

    However, if one considered the possible memories and synaptic
connections in the brain of a human with specified genes but with
arbitrary experiences, one would need even more digits to specify those
than just to specify the genes.  For example, there are about a hundred
billion neurons in the human brain, and each is linked to as many as ten
thousand others, giving perhaps a million billion ($10^{15}$)
connections.  If this were the total possible number of connections, and
we had to say which ones were actually connected, we would need a million
billion binary digits to specify this or roughly 30\% as many decimal
digits.  In this manner, there may be hundreds of thousands of times more
information in ``nurture'' (the conscious and unconscious memories
recorded in synaptic connections that one gains from experiences) than in
``nature'' (one's genes).  In a similar manner, a googol of information
about which Everett many-world one exists in would be far more than the
500 digits of information which reveals which kind of universe one is in.

    On the other hand, each of these 500 digits would likely have a far
larger effect on the nature of the universe than each of the googol of
digits specifying the particular universe, just as the information in
each gene for a human may exude more of an effect than the information
stored in a single unit of memory.  Thus the nature vs. nurture debate
does not end merely with the observation that there is generally far more
information in nurture than in nature.

    Finally, there is a second googol of information that specifies where
one exists within a particular Everett many world. This may be taken to
be analogous to which lifetime experience a particular human of
particular genes is experiencing.  If one takes a person who is conscious
for two-thirds of a 70-year lifetime, roughly 1.5 billion seconds, and
says that he or she has one experience each tenth of a second, then this
person would have about 15 billion experiences within his or her
lifetime.  This would require just 11 digits to specify which lifetime
experience a particular person is presently experiencing.  These 11
digits are enormously less than the googol of digits it might take to
specify where we are within our particular universe of a particular kind
that has undergone inflation.  But then, unlike the universe, we have not
exponentially inflated.

    The vast size of the entire multiverse makes it seem likely that
almost all possible human experiences would occur somewhere. Indeed, if
inflation is true, it seems likely that there are very many copies of us,
spread throughout our vast universe, that have exactly the same genes and
memories.

    This existence of an enormous number of humans or other intelligent
life forms would be true even if the probability per planet for
intelligent life and consciousness were very small.  In particular, we
would likely exist in all of our details even if the probability per
planet were so small that it is unlikely that there would be any other
intelligent life and consciousness on any of the other million billion
billion ($10^{24}$) or so planets that may exist within the part of our
universe that we can see.

    One implication is that we would have no real evidence to suggest
that the probability per planet is large enough for us to have contact
with other intelligent life.  We cannot rule out the possibility of
contact either, but the fact that the probabilities per planet could be
so much smaller than one in a million billion billion suggests to me that
it may well be just wishful thinking to suppose that we can ever be in
contact with extraterrestrial intelligence.

    Does this picture of a vast universe or multiverse make us
insignificant, since we make up such a tiny part of it?  I believe not.

    The mere size of something does not determine its importance. Perhaps
because of our past human history, when human physical strength (often
correlated with size) was important for obtaining food, we have a
tendency to admire physical size and strength in humans, for example,
referring to someone as a giant in his or her field.  But now, if we stop
to think about it, we realize that this is mainly just a metaphorical
expression, and importance really has very little to do with physical
size.

    For example, the author of Psalm 8, after implicitly recognizing that
a human is far smaller than the heavens, responded with a praise of human
glory:  ``For You have made him a little lower than the angels, and You
have crowned him with glory and honor.''

    Even on a purely physical or mechanical level, it seems very likely
that we are the most complex beings within our solar system, and perhaps
even within the entire part of the universe that we can see.

    Another issue arises with the view that the entire multiverse
contains not just the observed humans on earth, but also presumably huge
multitudes of other humans and other intelligent and conscious beings,
widely distributed across space in our universe and across different
universes and maybe even across different types of universes.  We might
fear that in this view we would lose significance because of our loss of
uniqueness.

    It is indeed human nature to appreciate being considered unique.  I
remember feeling flattered when John Wheeler once wrote about me, ``No
one else has his combination of talents.''  However, I quickly realized
that he could have written the same about any person on earth (at least
if the comparison were to other people on this same earth).  One might
fear that this distinction that each of us has in comparison with the
hundred billion or so humans that have lived on earth would be lost if we
extended our comparison to the vast reaches of space, where we are likely
not to be so unique.

    But on further reflection, I do not understand why being unique is so
important.  Even without considering his or her unique combination of
talents, each person on earth is important just for being human, despite
the fact that there have been nearly a hundred billion other humans
living on earth.  So why can't each person be important for being who he
or she is, even if there are googolplexes of copies spread over the
entire universe?

    The urge to achieve importance through uniqueness has led humans to
seek roles that are obviously unique within a society, such as being the
ruler.  But if we can realize that each of us is important, whether or
not he or she is unique in any particular way, then we can be happy with
our lives and fulfill our roles to show love to others who are also
equally important.


\begin{thebibliography}{99}

\bibitem{Bible} This and all other Scripture taken from the New King
James Version.  Copyright \copyright 1982 by Thomas Nelson, Inc.  Used by
permission.  All rights reserved.

\bibitem{Danielson} Dennis Richard Danielson, ed., {\em The Book of the
Cosmos:  Imagining the Universe from Heraclitus to Hawking} (Perseus
Publishing, Cambridge, Massachusetts, USA, 2000).

\bibitem{Hogg} Helen Sawyer Hogg, ``Le Gentil and the Transits of Venus,
1761 and 1769,'' {\it Journal of the Royal Astronomical Society of Canada}
{\bf 45}, 37-44, 89-92, 127-134 \& 173-178 (1951).

\bibitem{Spergel} D.~N.~Spergel {\it et al.} [WMAP Collaboration],
``Wilkinson Microwave Anisotropy Probe (WMAP) Three Year Results:
Implications for Cosmology,'' {\it Astrophysical Journal Supplement
Series} {\bf 170}, 377 (2007), astro-ph/0603449,
$<$http://arxiv.org/abs/astro-ph/0603449$>$.

\bibitem{Linde} Andrei D. Linde, {\em Particle Physics and Inflationary
Cosmology} (Harwood, Chur, Switzerland, 1990).

\bibitem{Guth} Alan Guth, {\em The Inflationary Universe: The Quest for a
New Theory of Cosmic Origins} (Addison-Wesley, Reading, Massachusetts,
USA, 1997).

\bibitem{Vilenkin} Alex Vilenkin, {\em Many Worlds in One: The Search for
Other Universes} (Hill \& Wang, New York, USA, 2006).

\bibitem{many-worlds} Bryce DeWitt and R. Neill Graham, eds, {\em The
Many-Worlds Interpretation of Quantum Mechanics} (Princeton University
Press, Princeton, New Jersey, USA, 1973).

\bibitem{Susskind} Leonard Susskind, {\em The Cosmic Landscape:  String
Theory and the Illusion of Intelligent Design} (Little, Brown and
Company, New York, 2006).

\bibitem{DouKac} Michael R.~Douglas and Shamit Kachru, ``Flux
Compactification,'' {\it Reviews of Modern Physics} {\bf 79}, 733-796
(2007), hep-th/0610102 $<$http://arxiv.org/abs/hep-th/0610102$>$.

\end{thebibliography}
\end{document}